\documentclass[11pt]{article} 
\usepackage{hyperref}
 
\pdfoutput=1 
 
\begin{document} 
 
\title{Sessile droplet evaporation on \\ superheated superhydrophobic surfaces} 
 
\author{Robb C. Hays, Julie Crockett, Daniel Maynes, Brent W. Webb \\ 
\\\vspace{6pt} Department of Mechanical Engineering, \\ Brigham Young University, Provo, UT 84602, USA} 
 
\maketitle 
 
\begin{abstract}
This fluid dynamics video depicts the evaporation of sessile water droplets placed on heated superhydrophobic (SH) surfaces of varying cavity fraction, $F_c$, and surface temperature, $T_s$, above the saturation temperature, $T_{sat}$. Images were captured at 10,000 FPS and are played back at 30 FPS in this video. Teflon-coated silicon surfaces of $F_c = 0,\,0.5,\,0.8,\,\mathrm{and}\,0.95$ were used for these experiments. $T_s$ ranging from $110\mathrm{^{\circ}C}$ to $210\mathrm{^{\circ}C}$ were studied. The video clips show how the boiling behavior of sessile droplets is altered with changes in surface microstructure. Quantitative results from heat transfer rate experiments conducted by the authors are briefly discussed near the end of the video.
\end{abstract}

\section{Introduction}

The hydrophobicity of a surface is the degree to which it repels water. Hydrophobicity can be quantified by the internal contact angle, $\theta$, a water droplet makes with the surface (Wenzel 1936). The theoretical upper limit on $\theta$ for a smooth surface is $120^{\circ}$. A superhydrophobic (SH) surface can therefore be defined as one which exhibits $\theta > 120^{\circ}$. Superhydrophobicity is achieved through a combination of microscale surface roughness and hydrophobic surface coating (Barthlott et al. 1997). SH surfaces can be characterized by their cavity fraction, $F_c$, defined as the ratio of projected unwetted area to total projected area of the surface. $F_c = 0$ for smooth surfaces and approaches unity as hydrophobicity is increased.

Droplets placed on heated smooth surfaces at temperature, $T_s$, above the saturation temperature, $T_{sat}$, experience varying degrees of nucleate boiling from the onset of nucleate boiling near $T_s - T_{sat} = 5\mathrm{^{\circ}C}$ to the Leidenfrost point near $T_s - T_{sat} = 120\mathrm{^{\circ}C}$. Heat transfer to droplets reaches a maximum at the critical heat flux point near $T_s - T_{sat} = 30\mathrm{^{\circ}C}$ (Incropera et al. 2007). Droplets placed on SH surfaces exhibit altered boiling behavior owing to their surface roughness.

\section{Methods}

This fluid dynamics video presents 10,000 FPS high speed video of sessile droplets undergoing evaporation on a smooth hydrophobic surface along with three rib-patterned SH surfaces of $F_c = 0.5,\,0.8,\,\mathrm{and}\,0.95$, respectively. Images were captured at $T_s$ ranging from $110\mathrm{^{\circ}C}$ to $210\mathrm{^{\circ}C}$. An SEM image of one of the SH surfaces used is included before the videos to show the surface roughness structure and scale. An image defining internal contact angle, $\theta$, is also included for reference. Droplet evaporation videos are then shown accompanied by explanatory remarks. Results of an experimental analysis conducted by the authors of heat transfer rate vs. excess temperature, $T_s - T_{sat}$, are included at the end of the video for interpretive purposes.

\section{References}

Barthlott, W., and Neinhuis, C., 1997. "Purity of the sacred lotus, or escape from contamination in biological surfaces." {\it Planta}, {\bf 202}(1), April, pp. 1--8.
\\[10pt]
Incropera, F., DeWitt, D., Bergman, T., and Lavine, A., 2007. {\it Fundamentals of Heat and Mass Transfer, 6th Ed.}. John Wiley \& Sons, Hoboken, NJ.
\\[10pt]
Wenzel, R., 1936. "Resistance of solid surfaces to wetting by water." {\it Industrial and Engineering Chemistry}, {\bf 28}(8), August, pp. 988--994.

\end{document}